\documentclass[conference]{IEEEtran}
\IEEEoverridecommandlockouts
\usepackage{cite}
\usepackage{amsmath,amssymb,amsfonts}
\usepackage{algorithmic}
\usepackage{graphicx}
\usepackage{textcomp}
\usepackage{xcolor}

\usepackage{listings}

\graphicspath{{figs/}{figures/}{pictures/}{images/}{./}} 

\usepackage{tabu}                      
\usepackage{booktabs}                  
\usepackage{lipsum}                    
\usepackage{mwe}                       

\usepackage{mathptmx}                  
\usepackage{url} 
\usepackage{amssymb}
\usepackage{multirow}
\usepackage{enumitem}
\usepackage[normalem]{ulem}

\usepackage{tikz}
\usepackage{xcolor}
\usepackage[scale=0.82]{FiraMono}

\definecolor{maroon}{RGB}{138, 16, 11}
\definecolor{gray}{RGB}{114, 97, 88}
\definecolor{gold}{RGB}{220, 202, 160}
\definecolor{red}{RGB}{179, 0, 0 }
\definecolor{green}{RGB}{93,105,74}

\newcommand*\rectRoundedGreen[1]{\tikz[baseline=(char.base)]{
            \node[shape=rectangle,rounded corners,fill=green, inner sep=2.5pt] (char) {\textcolor{white}{\normalfont\texttt{\large{#1}}}};}}

\newcommand*\rectRoundedRed[1]{\tikz[baseline=(char.base)]{
            \node[shape=rectangle,rounded corners,fill=red, inner sep=2.5pt] (char) {\textcolor{white}{\normalfont\texttt{\large{#1}}}};}}

\def\BibTeX{{\rm B\kern-.05em{\sc i\kern-.025em b}\kern-.08em
    T\kern-.1667em\lower.7ex\hbox{E}\kern-.125emX}}
\begin{document}

\title{ChatGPT in Data Visualization Education: A Student Perspective\\

}

\author{

\IEEEauthorblockN{Nam Wook Kim}
\IEEEauthorblockA{\textit{Computer Science} \\
\textit{Boston College}\\
Chestnut Hill, United States \\
nam.wook.kim@bc.edu}
\and
\IEEEauthorblockN{Hyung-Kwon Ko}
\IEEEauthorblockA{\textit{Computer Science} \\
\textit{KAIST}\\
Daejeon, Republic of Korea \\
hyungkwonko@gmail.com}
\and
\IEEEauthorblockN{Grace Myers}
\IEEEauthorblockA{\textit{Computer Science} \\
\textit{Boston College}\\
Chestnut Hill, United States \\
grace.myers@bc.edu}
\and
\IEEEauthorblockN{Benjamin Bach}
\IEEEauthorblockA{\textit{Bivwac} \\
\textit{INRIA}\\
Bordeaux, France \\
benjamin.bach@inria.fr}

}

\maketitle

\begin{abstract}
Unlike traditional educational chatbots that rely on pre-programmed responses, large-language model-driven chatbots, such as ChatGPT, demonstrate remarkable versatility to serve as a dynamic resource for addressing student needs from understanding advanced concepts to solving complex problems. This work explores the impact of such technology on student learning in an interdisciplinary, project-oriented data visualization course. Throughout the semester, students engaged with ChatGPT across four distinct projects, designing and implementing data visualizations using a variety of tools such as Tableau, D3, and Vega-lite. We collected conversation logs and reflection surveys after each assignment and conducted interviews with selected students to gain deeper insights into their experiences with ChatGPT. Our analysis examined the advantages and barriers of using ChatGPT, students' querying behavior, the types of assistance sought, and its impact on assignment outcomes and engagement. We discuss design considerations for an educational solution tailored for data visualization education, extending beyond ChatGPT's basic interface.

\end{abstract}


\section{Introduction}
Chatbots have been widely used in educational settings as basic interactive tools that assist in delivering curriculum content, facilitating practice exercises, and simulating simple tutor-student interactions~\cite{kuhail2023interacting}. These earlier chatbots were often rule-based systems that relied on scripted responses and could handle frequently asked questions or provide predefined explanations~\cite{cunningham2019review}. On the other hand, recent large language model (LLM)-driven chatbots such as ChatGPT have garnered significant attention across our society by their remarkable versatility in performing a wide array of tasks across different sectors. The potential implications for education are also profound, supporting more complex educational tasks such as drafting essays and solving intricate mathematical problems~\cite{kasneci2023chatgpt}. While there is growing interest and exploration surrounding the challenges and opportunities of these chatbots in education, research on their actual impact on student learning practices remains limited, particularly true for understanding how students utilize these tools in real-world, open-ended creative tasks. 

In this study, we explore the utility of ChatGPT for undergraduate students in a data visualization course by identifying advantages and potential barriers across four different assignments throughout the semester. We selected ChatGPT as a representative of LLM-based chatbots for our study due to its widespread use and high performance during the fall semester of 2023~\cite{ahmed2023chatgpt,borji2023battle}. Students used various tools, such as Tableau, D3.js, and Vega-lite, to design data visualizations and develop interactive dashboards and data-driven articles. They completed survey questionnaires and submitted conversation logs as part of their assignments. At the end of the semester, we conducted a brief final reflection survey and carried out follow-up interviews with selected students. We explore to answer the following research questions:

\begin{itemize}[noitemsep,topsep=0pt]
    \item What are the challenges of ChatGPT use in data visualization tasks? $\rightarrow$ We look into survey responses and interview results across different assignments.
    \item How do students' querying behaviors and the types of assistance they seek vary? $\rightarrow$ We analyze conversation logs and inspect the types of questions asked. 
    \item What is the impact of ChatGPT on their learning outcomes and engagement? $\rightarrow$ We examine how factors like conversation volume and attitude relate to course grades.
\end{itemize}

All student participants expressed positive feedback regarding the use of ChatGPT in the course. They reported that ChatGPT enhanced various aspects of their assignment experience by accelerating the completion process, improving the quality of their assignments, and boosting their confidence. When analyzing individual assignments, participants used ChatGPT more frequently for programming tasks, while finding it less beneficial for the Tableau assignment. Despite these advantages, the results also highlighted several usability and technical challenges associated with ChatGPT. These included difficulties in phrasing questions, the absence of context in responses, challenges with text-based interactions, and the impact of inaccurate responses.

Upon analyzing the conversation logs, we identified three high-level querying themes in order of frequency: solution-oriented queries (seeking a specific outcome with a clear goal, such as ``help me create HTML''), abstract and open-ended queries (exploring options without a definitive outcome, such as ``make my website prettier''), and interpersonal queries (expressing emotional feedback such as, ``margins not applying!''). We initially hypothesized that more determined and frequent use of ChatGPT might lead to better grades. However, when analyzing the relationship between the query themes and student grades, we did not observe any significant impacts. Similarly, our examination of the correlation among the number of interactions, query lengths, and grades also revealed no significant effects.

Based on the lessons learned from our study, we discuss design considerations for a future educational system tailored to data visualization education. These considerations include promoting design-oriented queries, supporting diverse inputs and outputs for fluid interactions, and evolving beyond merely being a tool for chart implementation. Our contribution lies in providing an improved understanding of how students use ChatGPT in a design-oriented, project-based course.

\section{Background \& Related Work}
\label{sec:related-work}
The integration of artificial intelligence in educational settings has evolved over the past decades, transitioning from rule-based tutoring systems that mimicked human-tutor interactions to more advanced systems employing machine learning and natural language processing~\cite{hwang2020vision,nkambou2010advances}. Examples include Cognitive Tutors, which uses model-tracing to provide step-by-step coaching~\cite{anderson1995cognitive}, AutoTutor, which engages students in conversational interactions \cite{graesser2004autotutor}, and ASSISTments, which blends tutoring assistance and assessment reporting~\cite{heffernan2014assistments}.

Intelligent tutors are used across various educational domains, serving roles such as subject tutors in science and mathematics, administrative assistants for course scheduling, and support for special education needs related to cultural diversity and accessibility~\cite{perez2020rediscovering,hwang2023review}. Several authoring tools have been developed to facilitate the creation of these intelligent tutors, including the Cognitive Tutor Authoring Tools~\cite{aleven2016example}, ASSISTments Builder~\cite{razzaq2009assistment}, and AutoTutor tools~\cite{nye2014autotutor}. More recently, general-purpose cloud-based chatbot platforms such as Dialogflow~\cite{dialogflow}, Wit.ai~\cite{witai}, and Rasa~\cite{rasa} have simplified the process of building custom conversational agents~\cite{cunningham2019review}. These chatbots adopt various roles, including teachers and learning companions, and exhibit diverse interaction styles ranging from user-driven to machine-driven conversations~\cite{kuhail2023interacting}.

Studies highlight several advantages of chatbots, including offering unrestricted access irrespective of time and location~\cite{hwang2023review}, fostering interactive and adaptive learning~\cite{vanlehn2011relative}, and alleviating the burden on teachers by managing routine inquiries~\cite{labadze2023role}. These tools are capable of assisting each student individually by adapting to various learning paces and providing tailored knowledge based on their cognitive levels~\cite{labadze2023role}. However, challenges have been noted, particularly the chatbots' inability to fully grasp context and nuances as humans do, exacerbated by issues including knowledge accuracy and system maintenance~\cite{perez2020rediscovering,okonkwo2021chatbots}. Additionally, ethical concerns such as transparency and privacy have been raised, especially as agents autonomously learn students’ personal information~\cite{zawacki2019systematic,okonkwo2021chatbots}.

Recently, chatbots based on large language models, notably ChatGPT, have garnered widespread attention across society. These chatbots, trained on extensive datasets from the web, display a high degree of versatility in performing diverse tasks across various disciplines, including education. This AI advancement has sparked numerous scholarly inquiries and investigations into its educational potential. Researchers discuss the potential benefits and threats based on prior literature~\cite{kasneci2023chatgpt,labadze2023role,kooli2023chatbots} and reactions on social media~\cite{tlili2023if,mogavi2024chatgpt}. Generally, they highlight similar advantages to previous generations, such as serving as personal learning aids or instructional assistant tools, while also illustrating more diverse scenarios due to the versatility of LLMs, such as general content generation and problem-solving~\cite{mogavi2024chatgpt,labadze2023role}. On the other hand, there is increased concern regarding potentially exacerbating drawbacks, such as biases and errors, as well as a growing reliance on AI~\cite{kooli2023chatbots,kasneci2023chatgpt}. Researchers argue that educators need to prepare for changes in educational practices, including developing new pedagogical approaches and addressing issues of academic integrity~\cite{becker2023programming,tlili2023if}.

Despite the relatively short history of LLM-based chatbots, several studies have explored the perspectives of teachers and students. Yu discusses potential coping strategies for teachers, including developing collaborative pedagogy with AI, nurturing responsible AI engagement, and cultivating emotional intelligence~\cite{yu2024application}. Herft offers guidance on various use cases for prompting, ranging from creating lesson plans and teaching strategies to generating formative assessment questions~\cite{herft2023teacher}. Others conducted interviews with educators to gauge their viewpoints, revealing an increasing awareness, positive and negative perceptions, and ethical concerns~\cite{ghimire2024generative,wang2023exploring}. Jeon and Lee identified complementary roles for ChatGPT and human teachers~\cite{jeon2023large}. Other researchers have examined ChatGPT’s capabilities in recommending teaching instructions~\cite{jordan2024need}, generating new assignments~\cite{joshievaluating}, and completing existing assignments~\cite{chen2023beyond, shen2024implications, ouh2023chatgpt, joshi2024chatgpt,wang2023exploring}.

Other studies have found that students generally hold positive attitudes and experiences with ChatGPT~\cite{rogers2024attitudes,chan2023students,budhiraja2024s}. Students use it for a variety of tasks including writing essays, understanding complex concepts, exploring alternative solutions, and enhancing programming skills~\cite{yilmaz2023augmented,budhiraja2024s}. Another set of studies explored how students use LLM-based code generators and assistants in practical courses, identifying various coding approaches and usability considerations~\cite{kazemitabaar2024codeaid,liu2024teaching,rasnayaka2024empirical,kazemitabaar2023novices}. In controlled experiments, Vaithilingam found that using Copilot led to more frequent failures compared to using Intellisense~\cite{vaithilingam2022expectation}, while Mousa and Veilleux discovered that students who did not use ChatGPT in Mathematics performed better than those who did~\cite{mousa2024chatgpt}. Hou et al.'s study indicates that Gen AI has not yet surpassed traditional help resources, and that help-seeking with Gen AI is a skill that requires development~\cite{hou2024effects}.

Our research contributes to the rapidly expanding body of research on the impact of LLMs in education. Specifically, our study focuses on the student experience and perspectives regarding the use of ChatGPT in an interdisciplinary data visualization course that requires both design and implementation skills.

\section{Methodology}
This section describes the procedures, participants, and tasks used to address the research goals outlined earlier.

\subsection{Procedures}

The study was conducted as part of the visualization course in the fall semester of 2023, which is offered as an elective for upper-class undergraduate students. Students were informed of the opportunity to participate in the study at the beginning of the course, and they signed informed consent forms describing the experiment details.

Participants completed four assignments during the course (see Section \ref{sec:tasks} Tasks). For each assignment, students were instructed to log their conversations with ChatGPT using a dedicated chat and asked to submit the log in the markdown format using the \textit{Save ChatGPT} Chrome extension. After each assignment, they were required to fill out a reflection survey regarding their experience. 

Upon course completion, students completed a final post-course survey. Selected students also participated in follow-up Zoom interviews, which required separate consent forms. Audio recordings were saved for later analysis. Interviewees were asked about their ChatGPT experience, with questions based on their survey responses. The interviews took place between December 15, 2023, and January 4, 2024. The interviews varied in length, ranging from 17 to 38 minutes, with a median length of 26 minutes.

Participation in the study was voluntary. Students received 0.5\% extra credit per assignment, up to 2\% total.  Students who chose not to participate could earn equivalent credit by completing a separate assignment where they were tasked with creating an \textit{original} visualization. Interviewees received a \$25 Amazon gift card as compensation.

\subsection{Participants}

\subsubsection{Survey participants}
Out of the 39 students enrolled in the course, 34 consented to participate in the study, and 26 engaged in all subsequent surveys and submitted their ChatGPT logs. Our survey analysis considers only these 26 student participants.

The participants' prior exposure to ChatGPT varied, with all participants having heard of or interacted with ChatGPT before the course. Additionally, 13 participants had exclusively used ChatGPT, 11 had used other chatbots or assistants and preferred ChatGPT, and 2 had no preference. A majority reported using ChatGPT weekly (15/26), while 6 used it daily, 3 monthly, and 2 rarely. The primary reasons for using ChatGPT were diverse, including academic assistance (22/26), personal curiosity (20/26), professional work (18/26), and entertainment or casual conversation (18/26). In terms of benefits, clarification of complex topics (22), quick answers to questions (21/26), ideas or inspiration (21/26), and fun \& entertainment (11/26) were most cited. Among various concerns, response accuracy was a primary issue (22/26). 

Regarding data visualization experience, more than half of the participants (15/26) had not been introduced to or worked on data visualization prior to this course. Most identified themselves as beginners in data visualization (19/26), with 6 at an intermediate level, and only 1 participant self-identified as advanced. The tools used for visualization varied, with Excel or Google Sheets being the most common (23/26), followed by R or Python for statistical visualization (16/26), and a few used more specialized software like PowerBI (3/26), Tableau (2/26), and Vega-lite (2/26). The majority had engaged in creating visualizations for academic assignments or projects (20/26), while 15 had done so for work-related purposes, three had no prior experience in creating visualizations before this course, and three had created visualizations for personal projects or for fun.

\subsubsection{Interview Participants}
Out of 26 study participants, we selected 14 for follow-up interviews. In choosing these participants, we aimed to balance the students' grades and the amount of ChatGPT usage. Table \ref{tbl:interview-participants} shows the demographics of the interview participants.

\begin{table*}[]
\centering
\begin{tabular}{rllrlrlrlrlll}
\hline
\multicolumn{1}{l}{\multirow{2}{*}{PID}} & \multirow{2}{*}{Class} & \multirow{2}{*}{Major}     & \multicolumn{2}{l}{Assignment 1}     & \multicolumn{2}{l}{Assignment 2}     & \multicolumn{2}{l}{Assignment 3}     & \multicolumn{2}{l}{Assignment 4}     & \multirow{2}{*}{Exam} & \multirow{2}{*}{\parbox{0.6cm}{Final \\ Grade} } \\ \cline{4-11}
\multicolumn{1}{l}{}                     & & & \multicolumn{1}{l}{Grade} & ChatSize & \multicolumn{1}{l}{Grade} & ChatSize & \multicolumn{1}{l}{Grade} & ChatSize & \multicolumn{1}{l}{Grade} & ChatSize &                       &                              \\ \hline
P1 & Senior & Computer Science & 0.99 & 90 & 1.00 & 26 & 0.98 & 260 & 0.98 & 248 & 0.85 & 0.99\\
P2 & Senior & Finance & 0.99 & 34 & 0.90 & 50 & 1.00 & 86 & 0.89 & 18 & 0.89 & 0.96\\
P3 & Senior & Computer Science & 1.00 & 16 & 0.98 & 26 & 1.00 & 34 & 1.00 & 58 & 0.97 & 1.01\\
P4 & Senior & Computer Science & 1.00 & 110 & 1.00 & 270 & 1.00 & 46 & 1.00 & 50 & 0.93 & 1.01\\
P5 & Senior & Computer Science & 1.00 & 26 & 1.00 & 16 & 1.00 & 58 & 1.00 & 54 & 0.89 & 1.00\\
P6 & Senior & Finance/Business Analytics & 1.00 & 20 & 1.00 & 26 & 0.90 & 236 & 1.00 & 84 & 0.91 & 0.99\\
P7 & Senior & Business Analytics & 0.87 & 200 & 0.80 & 86 & 0.80 & 76 & 0.69 & 56 & 0.91 & 0.83\\
P8 & Senior & Computer Science & 0.94 & 18 & 0.86 & 10 & 0.82 & 36 & 0.96 & 34 & 0.71 & 0.91\\
P9 & Senior & Computer Science/Finance & 1.00 & 132 & 1.00 & 116 & 1.00 & 200 & 1.00 & 274 & 0.98 & 1.02\\
P10 & Senior & Computer Science & 1.00 & 196 & 1.00 & 70 & 1.00 & 152 & 1.00 & 186 & 0.91 & 1.01\\
P11 & Senior & Mathematics & 0.80 & 22 & 0.99 & 54 & 0.74 & 32 & 1.00 & 16 & 0.88 & 0.93\\
P12 & Senior & Computer Science & 0.99 & 162 & 1.00 & 46 & 0.98 & 194 & 1.00 & 178 & 0.94 & 1.01\\
P13 & Senior & Finance/Business Analytics & 1.00 & 26 & 1.00 & 68 & 1.00 & 24 & 1.00 & 62 & 0.92 & 1.01\\
P14 & Junior & Computer Science & 1.00 & 30 & 0.92 & 40 & 0.90 & 58 & 1.00 & 100 & 0.91 & 0.98\\
\hline
\end{tabular}
\vspace{0.2cm}
\caption{Demographics of interview participants: grades are in percentages, ChatSize represents the number of turn-taking interactions with ChatGPT, exam scores are from an in-class quiz on lecture material, and final grades include extra credit.}
\label{tbl:interview-participants}

\vspace{-0.5cm}
\end{table*}

\subsection{Task Assignments}
\label{sec:tasks}
The visualization course had four distinctive assignments described below. 

\subsubsection{Design \& Redesign}
In the first assignment, students are tasked with performing two main activities: 1) creating a static data visualization using a student enrollment dataset, and 2) redesigning Florence Nightingale's Coxcomb chart, accompanied by a well-articulated rationale. They need to integrate these components into a website. The objective is to practice skills in creating visualization designs and to enhance the students' abilities to articulate their design choices.

\subsubsection{Exploration and exposition with Tableau}
In the second assignment, students are tasked with exploring datasets and constructing dashboards or story points using Tableau. They have been provided with two choices of real-world datasets: a household survey on internet access and a tree planting history dataset, both from local cities. They are required to integrate their Tableau outcomes into a website. The objective is to practice exploratory analysis and develop visual data storytelling skills.

\subsubsection{Interactive Dashboard using D3}

In the third assignment, students are tasked with constructing multiple coordinated interactive visualizations using D3. They were provided with the same datasets as the second assignment. Technical requirements were in place including at least three visualizations with at least two different chart types and coordinating interactions. The goal is to practice web-based interactive visualization implementation skills.

\subsubsection{Data-Driven Article using D3 \& Vega-lite}

In the final assignment, students are assigned the task of creating a data-driven article that integrates narrative text and data visualizations using the same provided datasets. The primary technical requirements stipulate that at least one chart must be created using Vega-Lite, feature in-chart annotations, and be interactively coordinated with the text~\cite{barry2014visualizing}. The goal is to practice the synthesis of text and visuals for engaging data storytelling.

\section{Data \& Analysis Methods}
Our final dataset comprises survey responses, conversational logs, and end-of-course surveys from 26 participants, plus interview transcripts from 14 of these participants.

\subsection{Survey \& Interview Analysis}

For analyzing surveys, we primarily used frequency analysis to compute percentage distributions of survey responses. For interviews, two researchers conducted open coding of the transcripts. They organized quotes in a spreadsheet based on themes and categories, initially derived from interview questions (e.g., strengths, weaknesses, feedback, and improvements). Through multiple collaborative meetings, they refined and recategorized responses into emerging themes, resolving conflicts as they arose. The final themes, categorized as advantages (\rectRoundedGreen{+}) or barriers (\rectRoundedRed{-}), include coding assistance, accessibility, serendipitous learning, and task-dependent effectiveness. These themes are directly presented in Section \ref{sec:interview-result}, with participant IDs for transparency.




\subsection{Conversational Log Analysis}

We gathered 3,773 user queries from 26 participants. Since the original query data is often too long, we manually processed user queries by replacing parts that are not essential for understanding the overall themes of the query dataset. Specifically, we replaced the Javascript, Python, HTML code into \texttt{$<$CODE$>$}, csv or tsv format data into \texttt{$<$DATA$>$}, https link to \texttt{$<$LINK$>$}, evaluation rubrics and excessive textual explanation into \texttt{$<$INFO$>$}, and error message to \texttt{$<$ERROR\_MSG$>$}.

To perform thematic analysis more efficiently, we employed an automatic qualitative coding approach leveraging large-language models, following the recent approach ~\cite{hamalainen2023evaluating}. The method simulates the overall process of humans doing the thematic analysis. First, we generated representative codes for each query by feeding the following prompt to GPT-4-turbo by OpenAI:

\lstset{basicstyle=\small\ttfamily,
        breaklines=true,
        literate={\{}{{\{}}1
             {\}}{{\}}}1,
}
\begin{lstlisting}
Let's perform a thematic analysis in the field of human-computer interaction. The given query is used in a conversation between a user and ChatGPT regarding design knowledge and implementation skills in a data visualization course. Extract the characteristics of high-level intention of user query using a few words. The total number is at most five, and each of them is separated by semicolons. Do not add numbering or any explanations.

Query: {query}

##
; ; ; ;
\end{lstlisting}

We limited the number of codes to a maximum of five per query to prevent overfitting. After manually examining the codes for quality assurance and filtering out irrelevant and erroneous entries, we reduced the initial 18,865 codes to 16,017 valid codes. We then consolidated overlapping codes, resulting in 4,043 unique codes.
To capture broader themes from these codes, we applied a clustering algorithm. First, we computed semantic embedding vectors of the codes using OpenAI embeddings~\footnote{https://platform.openai.com/docs/models/embeddings}, and then reduced the dimensionality of these vectors using UMAP. Next, we applied HDBSCAN to combine similar codes into semantically relevant topics.
We evaluated topic prevalence by counting the frequencies of the code clusters (i.e., the number of user queries assigned at least one code from each cluster). Through manual examination, we further consolidated these clusters into three higher-level themes, as described in Section \ref{sec:log-analysis-result}. We also examined the distribution of these themes among different users and assignments to gain insights into their relative importance and occurrence patterns.




\subsection{Integrated Analysis}
To understand how students' experiences with and attitudes toward ChatGPT impact their course outcomes, we analyzed the relationship between ChatGPT conversation usage, survey responses, and course grades. In other words, we conducted statistical tests to determine whether various factors, such as the number of interactions with ChatGPT, the lengths of queries, and the themes of queries, correlate with students' perceptions and their academic performance.

\section{Results}

\subsection{Final Course Survey Result}

\begin{figure*}
  \centering
  \includegraphics[width=0.90\textwidth]{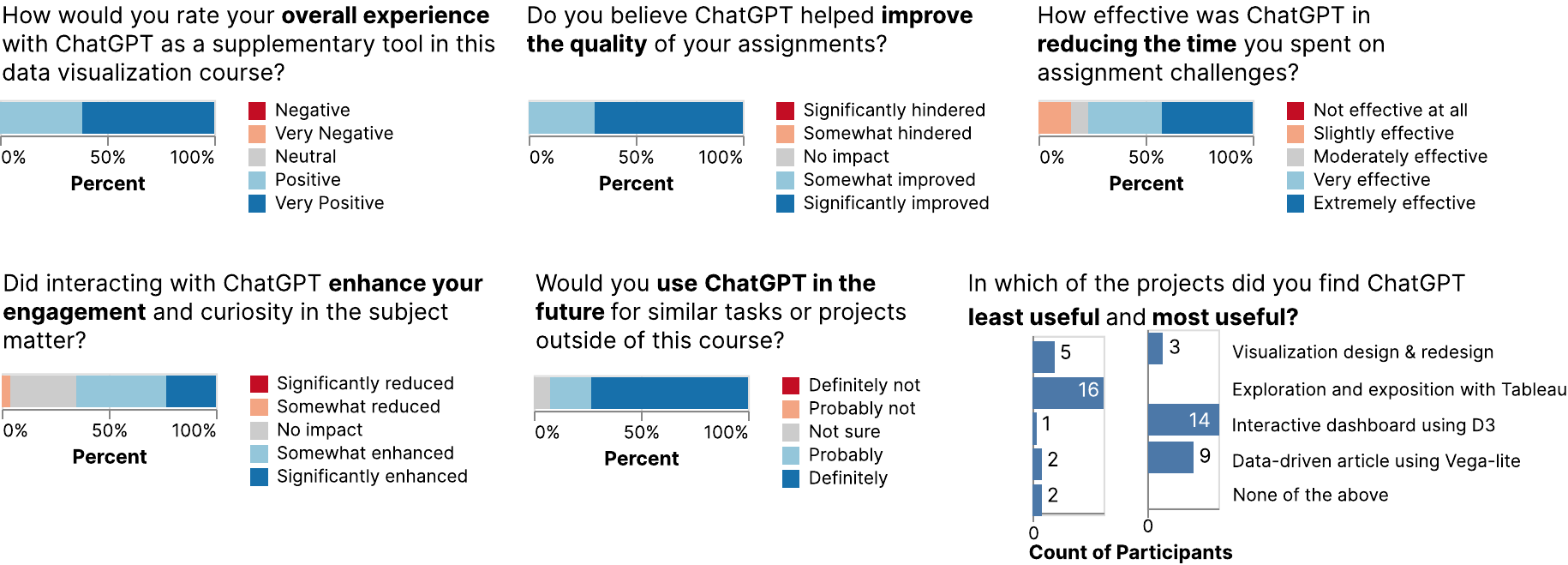}
  \caption{Survey responses from 26 participants on ChatGPT's impact in a data visualization course, encompassing overall experience, assignment quality, engagement, future use intention, and effectiveness in time management across various projects.}
  \label{fig:final-course-survey}
  
\end{figure*}

Figure \ref{fig:final-course-survey} presents the collective sentiment of students regarding the use of ChatGPT in the data visualization course. The overall experience was reported as favorable, with 69\% of students rating their experience as very positive and 31\% as positive. Regarding the quality of assignments, 69\% believed that ChatGPT had significantly improved the quality, while 31\% felt it somewhat improved their work. In terms of efficiency, 42\% of students found ChatGPT to be very effective, and 35\% found it moderately effective at reducing time spent on assignment challenges. ChatGPT also seemed to positively affect student engagement, with 31\% reporting a significant enhancement in their interest and curiosity in the subject matter, and 42\% noticing some enhancement. Future use of ChatGPT appears to be highly anticipated, with 92\% of students indicating they would probably or definitely use ChatGPT for similar tasks or projects outside of this course. 

ChatGPT was found most valuable for the ``Interactive dashboard using D3'' project (54\%), while it was deemed least useful for ``Exploration and exposition with Tableau'' (62\%).  Key strengths of ChatGPT identified by participants included its assistance in ``Code Syntax and Debugging'' (89\%), ``Being Available 24/7'' (85\%), and its ability to ``Retrieve Information Quickly'' (65\%). On the other hand, notable challenges faced by users were the ``Inaccuracy of Information'' (78\%), the tendency to develop an "Over-reliance on AI for Problem-Solving" (59\%), and the ``Difficulty of Phrasing Questions'' for ChatGPT (44\%). These findings corroborate prior survey results that report students' use of ChatGPT, both in general and in programming courses.~\cite{rogers2024attitudes,chan2023students,budhiraja2024s,yilmaz2023augmented}.
 
\subsection{Intermediate Survey Results}

\begin{figure*}
  \includegraphics[width=\textwidth]{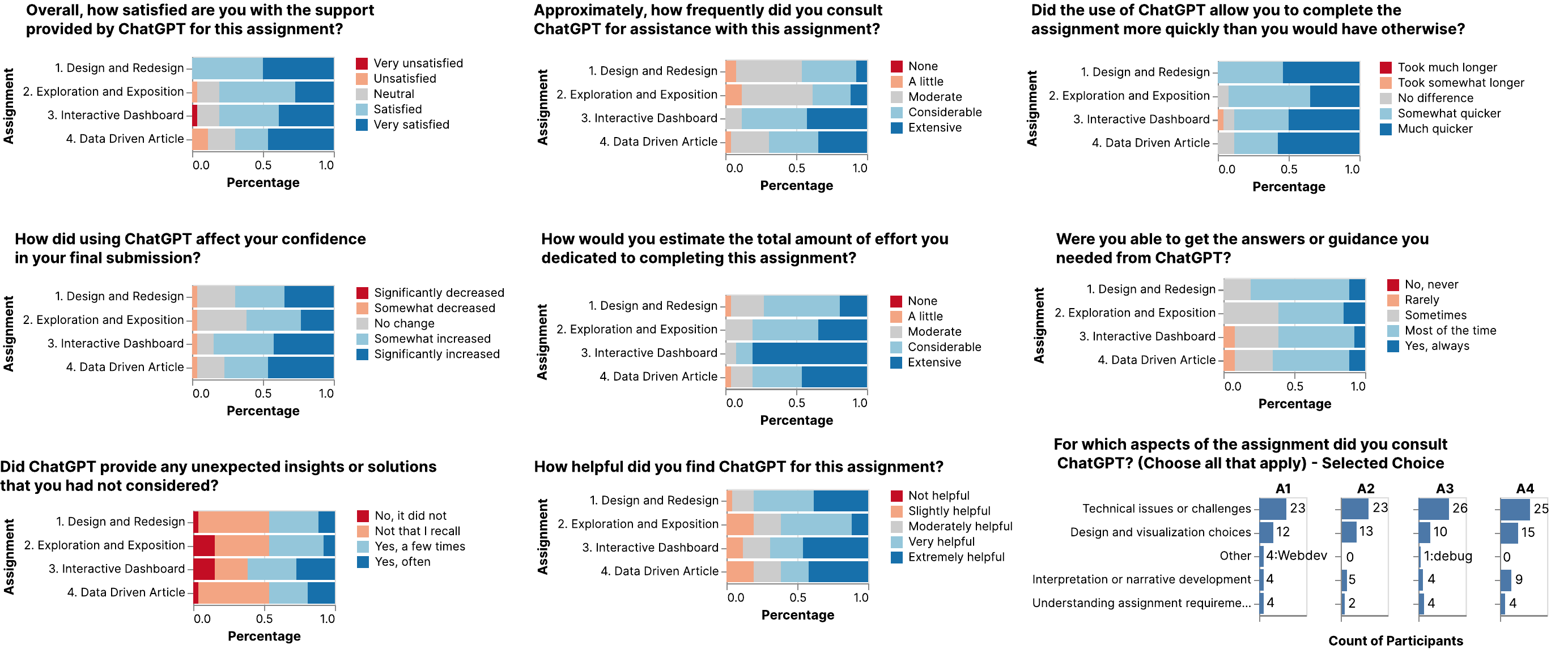}
  \caption{Intermediate survey results for individual assignments, illustrating satisfaction with support, confidence levels, frequency and helpfulness of ChatGPT interaction, speed of assignment completion, and areas of consultation within specific projects.}
  
  \label{fig:intermediate-survey-results}
\end{figure*}

Figure \ref{fig:intermediate-survey-results} presents an overview of the intermediate surveys conducted after each of the four assignments. 

Satisfaction with ChatGPT’s assistance across assignments showed a generally positive trend ($\ge$69\% satisfied). When assessing the influence of ChatGPT on participants' confidence, there was a noticeable trend of increased confidence as well ($\ge$61\% increased). The use of ChatGPT sped up assignment completion, with many participants noting they finished tasks quicker than they otherwise would have ($\ge$89\%). Generally, participants recognized the tool as helpful across most assignments, yet the sentiment was somewhat muted for the Exploration and Exposition using Tableau (see Figure \ref{fig:intermediate-survey-results}). Participants tended to consult ChatGPT more frequently for third and fourth assignments compared to first and second assignments ($\chi^2(12) = 21.50$, $p = 0.011$) 

Several other questions indicated less pronounced enthusiasm while still positive. For instance, when asked if ChatGPT consistently provided the needed answers or guidance, the most affirmative response ("Yes, always") was less prevalent compared to other options. Similarly, responses to whether ChatGPT offered unexpected insights were more tempered; the least affirmative response ("No, it did not") was more pronounced compared to other options. When asked about the aspects of the assignment they consulted on, most participants selected technical issues or challenges, followed by design and visualization choices (see Figure \ref{fig:intermediate-survey-results}).

In response to the question ``What features or capabilities do you wish ChatGPT had to better assist you with'', students suggested various ideas. 
For instance, many comments emphasized the need for ChatGPT to generate images or visualize data, reflecting their struggle with the textual limitations of ChatGPT when addressing inherently visual tasks like creating charts or designing websites. Another prominent trend was the call for enhanced coding assistance, particularly for languages such as HTML, CSS, JavaScript, and D3.js. Suggestions included the ability to render outcomes or accept various file types, including code, images, and data. Additionally, others recommended improvements in memory and context retention to enhance conversational interactions (e.g., ``sometimes it would undo a previous instruction I had given it'').

In ``Other Comments'', students provided additional feedback beyond the questions asked in the survey. Aligned with the survey questionnaire result (Figure \ref{fig:intermediate-survey-results}), several students pointed out ChatGPT's limitations when working with specific tools like Tableau (e.g., ``Can better assist me with using Tableau''). Moreover, some students also indicated that they began to rely less on it for complex problem-solving, as its effectiveness decreased with more complex or longer code snippets (e.g., ``when the task is complex, then ChatGPT will give wrong answers'').

\subsection{Post-Course Interview Results}
\label{sec:interview-result}

We categorize findings as advantages or barriers, but their impact varies among participants. Some students may overcome barriers that others find challenging. We report the main findings along with these exceptional cases as examples.

\subsubsection{\protect\rectRoundedGreen{+} Coding Assistance as a Primary Benefit}

The positive experiences with ChatGPT in the course primarily revolved around its usefulness in coding sections of the assignments, as noted by all participants. It helped fill the knowledge gap for those new or unfamiliar with web development (P1, P4) with P4 saying that ``[professor] didn't teach much about, like HTML, CSS, and JavaScript [...] I don't have any previous experience with those tools. [...] ChatGPT really helped me a lot.'' Others described various benefits including debugging \& error resolution (P1, P5, P6, P11, P12, P14), understanding advanced programming concepts (P1, P5, P7, P8, P14), generating starter code like templates \& outlines (P4, P5, P6, P8, P12), and boosting confidence in programming (P1, P5, P7, P9, P12). 

For example, P6 shared, ``If I copy and paste my code into the chat and ask, [...] it's able to recognize what's going on in my code [...]'', while P1 commented ``ChatGPT was sort of there as a buffer to help in my competence. Also, being a woman in STEM, you know, the confidence, it's not always there.'' Other participants were initially skeptical but were pleasantly surprised by ChatGPT's capabilities. P4 stated, ``I didn't expect that it can help me this much on the homework [...] ChatGPT just basically completely gave me the code.'' Similarly, P5 said, ``[...] creating a whole website only using ChatGPT like not even thinking it through at all. And I was just like, mind blown that you could do that.''

\subsubsection{\protect\rectRoundedRed{-} Limited Use for Visualization Design Guidance}

Although rubrics were in place for evaluating the quality of visualization designs, students did not utilize ChatGPT as much for design assistance (P2, P6, P10-P12, P14). For instance, P10 said ``I think for a lot of my projects, the main reason I use ChatGPT was to understand the code better and to have to be able to implement the code from class into our projects'', while P14 shared ``But anything we kind of learned in class and anything about design, or anything regarding creativity and stuff like that, I just revisited lecture slides for that.''

Although limited compared to what they learned in classes (e.g., choosing effective encoding choices based on perception, tasks, and data types), participants mentioned several basic use cases for seeking design knowledge. For instance, they mentioned chart suggestions as a useful feature (P3, P5 P7), such as showing a snapshot of the data and asking what chart would fit the data (P5, P7) or asking for an alternative representation to avoid the overuse of a bar chart (P9). Others also used ChatGPT to justify their design choices and identify the pros and cons of a chart (P3)---which was part of the first assignment requirement---asking for feedback for color schemes (P5, P7), and craft text narratives given charts (P3, P13). P3 gave a reason for the use case of design rationale generation, saying ``I could think of some of my own, but I thought ChatGPT articulated them better because it knew more like design choice words than I did at the time.''

\subsubsection{\protect\rectRoundedGreen{+} Improved Task Efficiency}

 All participants praised ChatGPT for its efficiency in saving time and aiding in project completion (P1, P2, P5-P10, P12, P14). Accelerated problem solving such as with automatic code generation and error detection was frequently mentioned as a contributing factor (P1, P6, P9, P14). For instance, P1 mentioned, ``it definitely was useful in doing the projects faster... I could just ask it a direct question about how would I change the background color of x, y, and z to be blue, and it would sort of give me that exact line.'' Some participants mentioned ChatGPT assisted their learning process (P3, P5, P7), as P7 said ``ChatGPT can kind of hold your hand through the process,'' The boosted productivity via ChatGPT often helped students focus on more important aspects of their projects including user experience and visualizations rather than syntax and debugging (P9, P12), or making the project better (P11, P12) or possible otherwise nearly impossible (P5, P10).
 
Similarly, ChatGPT's diverse knowledge base and ability to provide quick information retrieval were noted as strengths that enhanced the learning experience (P3, P6-P9, P11). For instance, participants contrasted ChatGPT with a search engine, with P11 saying ``Google, [...] gives you [...] more websites to look at. So that means you'll have to look through each website and kind of figure out oh, is this what I'm looking for [...], while P7 added ``ChatGPT does an incredible job of just boiling it down to some fundamental thing. You can just like in five bullet points, explain how to do this or something like that.'' P9 reinforced this perspective by describing ChatGPT as ``a base knowledge place where I can go to ask random questions or get quick, quick information before I kind of do a deeper dive on things,'' highlighting the tool's role as a starting point for broader research.

\subsubsection{\protect\rectRoundedGreen{+} Offering On-demand Accessibility}
Firstly, participants widely acknowledged ChatGPT's accessibility without physical and temporal constraints as a significant strength (P1, P5-P7, P11, P13, P14), which was particularly beneficial when working on assignments outside of typical office hours or when immediate assistance was needed. For example, P7 mentioned, ``being able to do it anytime anywhere, was super helpful...even just during the day, when you don't want to have to wait an hour and a half to figure out some little troubleshooting.'' Similarly, P13 emphasized the flexibility offered by ChatGPT, stating, ``I can arrange my time more flexibly so that I don't need to get it on before a specific office hour.'' This round-the-clock availability was contrasted with the limited accessibility of human instructors or teaching assistants (P7, P14).

\subsubsection{\protect\rectRoundedGreen{+} Catalyst for Initiating the Learning Process}

Many participants noted that ChatGPT was instrumental in helping them begin their assignments (P1, P3, P10, P13). For instance, students like P1 and P3 found ChatGPT helpful in ``getting started on assignments that seemed daunting'' by providing ``a framework'' for coding tasks. Similarly, P7 noted that it was ``extremely helpful to kind of get off the ground and at least just get some form of presentable kind of minimum viable product going.'' P1 added, ``even if it didn't give me exact answers, it could at least stimulate sort of the thoughts and thinking behind what might work.'' Similarly, P10 highlighted ChatGPT's role in providing guidance rather than exact answers, which helped them figure out solutions independently, `` [...] it would give me a basic skeleton of the code, and then I could just like, figure out the rest of myself, which I really liked.''

\subsubsection{\protect\rectRoundedGreen{+} Serendipitous Learning and Creative Exploration}

Some participants appreciated unexpected opportunities through their interaction with ChatGPT including unanticipated insights, problem-solving methods, and understanding of new concepts (P1, P3, P5, P7, P8, P14). They mentioned that ChatGPT guided them in exploring new stylistic aspects of their visualizations using HTML and CSS (P1, P3, P14) with P1 appreciating that such exploration made their projects ``more user-friendly and exciting''. Several participants noted that their work with ChatGPT led them to extend their technical skillset beyond their initial knowledge (P7, P14). For instance, P7's engagement with ChatGPT encouraged exploration of new capabilities of SVG and D3, while P14 discovered through ChatGPT a ``more robust, kind of more practiced way of like, creating websites, [...] more standard for the field.'' Other participants reflected on how their interaction with ChatGPT led to the discovery of new data visualization methods that they hadn't considered before (P3, P8). For example, P3's experience with ChatGPT led to the creation of a new chart based on tree sizes. This chart presented a novel way to visualize data that wasn't initially apparent. The participant described this as a breakthrough moment in the course. Similarly, P8 mentioned that ChatGPT helped them consider different chart types like heat maps and tree maps, which they had not deeply considered before.

\subsubsection{\protect\rectRoundedRed{-} Cost of Inaccurate and Misleading Responses}

Most participants experienced the occurrence of mistakes in the code generated by ChatGPT or response outputs that did not meet their expectations (P1-P3, P6, P11, P13, P14). For instance, P2 expressed frustration, stating, ``it saved me a good amount of time writing all the code. But sometimes it would make a little mistake, which would be annoying, because now I'd have to like, a lot of effort to fix it.'' Often participants observed ChatGPT pretend to understand user questions but keep generating incorrect responses, increasing the level of frustration (P1, P3, P13), as P13 described, ``[...] it cannot solve my problems, but it still pretends that I can, I will still believe it. And that just makes my life harder.'' Oftentimes, participants ended up asking repetitive follow-up questions (P6, P11) as P11 shared ``going into an endless loop with ChatGPT telling it that it's wrong. And then having to like, in the end, figure stuff out myself anyway.''

Inaccurate responses necessitated extra verification efforts, which ironically pushed some participants towards traditional learning resources (P3, P14), or affected the overall quality of work (P11, P14). P3 said ``Using ChatGPT would never be my go-to for getting interested in something after the course [...] I would definitely look to YouTube or anything else [...] like written by a human being that's knowledgeable in it [...]'', while P14 recalls a specific issue where ChatGPT did not identify a single erroneous line of code, while P11 describes abandoning a visualization approach due to persistent errors in ChatGPT’s responses. Some participants mentioned that it requires a certain level of prior knowledge to verify ChatGPT's responses and use it more effectively (P5, P7, P10). 

\subsection{\protect\rectRoundedRed{-} Difficulty of Phrasing Questions}

Another issue was the difficulty in phrasing questions and the misinterpretation of questions by ChatGPT (P1, P4, P5, P7-P10, P12, P13) Participants found that they had to be extremely specific to avoid vague answers (P4, P5, P13) or often they could not fathom what went wrong in their questions (P1), while others suggested that the ability to effectively interact with ChatGPT---or to ask effective questions---is contingent upon their existing knowledge base (P7, P12). A common coping strategy was to break down the problem into steps (P4, P11), as P11 noted ``Sometimes I tried to, like, talk through it in steps, because I figured out that that works a lot better than telling GPT to do everything at once.'' Others mentioned that their questioning skills got better over time (P5, P12, P13).

\subsubsection{\protect\rectRoundedRed{-} Lack of Contextual and Nuanced Understanding}

Participants also expressed frustration with ChatGPT's memory limitations and its impact on the effective understanding of user questions (P1, P5, P6, P14, P13). P5 shared, ``Sometimes [...] it would just like, forget where I was heading with it,'' indicating a challenge in maintaining context over extended interactions. This issue was echoed by P14 who noted that ChatGPT ``just kept trying to add code if there was an error, but it wouldn't really look at the existing code and try to delete things,'' suggesting a limitation in ChatGPT's ability to revise and refine its solutions. Similarly, another issue mentioned was ChatGPT's tendency to provide the same non-working solutions repeatedly (P7, P13). P7 described a situation where ``it would try to come back with code...sometimes it would even be the same exact code,'' indicating a lack of adaptability in the AI's responses. Participants sometimes had to start over with new sessions or repeatedly recontextualize their issues to get useful responses from ChatGPT (P8, P11).

\subsubsection{\protect\rectRoundedRed{-} Challenges with Text-Oriented conversation}

Several participants suggested that the ability for ChatGPT to process images or screenshots would be beneficial (P5, P11) with P11 stating, ``If ChatGPT was able to, like, read pictures or like images, it would be able to better understand what somebody is trying to communicate.'' Others also mentioned  the ability to input datasets and see the outputs of code (P1, P5, P14), with P1 suggesting ``if you could actually input like a dataset [...] see its outputs, and maybe if it displayed the graphs that it was creating, that might be more useful [...]''

\subsubsection{\protect\rectRoundedRed{-} Task-Dependent Effectiveness}

Another common trend was that ChatGPT was perceived as less effective for generating complex visualizations or interactions (P3, P6, P8, P12, P13). For instance, P6 mentioned ``I tried using ChatGPT for brushing, like a more advanced interaction, and it wasn't really able to help me with that.'', while P3 shared, ``[...] it's really good for these simple tasks. But if I wanted to do something like a complex, creative, interesting visualization, I couldn't use it for something like that.'' On the other hand, P2 and P14 mentioned that ChatGPT's knowledge base is limited for specific libraries including D3 and Vega-lite.

Similarly, participants indicated variable usage across assignments. For instance, P10 said, ``I think I use it a lot less in the first two assignments, just because those are more of like, using our creative abilities [...] I felt like I could take my own creative liberties with that.'' Many participants remarked on ChatGPT's limited effectiveness with Tableau projects (P1-P10, P12-14), often because the software did not require extensive coding or because Tableau's own resources were more straightforward.

\subsubsection{\protect\rectRoundedRed{-} Worries About Developing Dependence and Hindering Learning}

Several participants expressed worries about becoming overly dependent on ChatGPT (P1, P3, P4, P8, P11-P13). P1 cautioned about the ``over-reliance on AI,'' highlighting a tendency to ``let it do it for you''. P6 and P7 also mirrored these concerns, indicating a habitual turn to ChatGPT as the first response to coding problems. P14 felt that ChatGPT was ``most helpful if I already understood the concept, but just didn't really want to spend the time implementing it myself.'', implying a trade-off between efficiency and educational depth.

Participants were concerned such dependency potentially harms their learning.  P3 mentioned, ``I maybe knew how to [...] write the code in because I could figure out where to put my own information. But [...] I wouldn't actually know why I'm [...] choosing this specific kind of function.'' Similarly, P4 also mentioned, ``It's more like, accomplish assignment instead of like, [...] helped me understand how it works. Because I really don't know how it works.'' Likewise, P12 candidly stated that ``I don't think [ChatGPT] really helped with understanding class concepts'', while P7 similarly noted that ChatGPT ``enabled you to kind of bypass the learning process.'' 

\subsubsection{\protect\rectRoundedRed{-} Marginal Role in Engagement \& Curiosity Enhancement}

Many participants felt ChatGPT had little to no impact on their curiosity beyond the immediate classroom tasks. P2 and P6 noted a more utilitarian use of ChatGPT to save time, without it leading to further interest. P3, P4, P8, P11, P12, and P13 were among those who did not feel that ChatGPT stimulated their interest outside of the coursework, often citing a preference for traditional research methods or expressing a sense of contentment with using ChatGPT solely as a problem-solving tool. On the other hand, a subset of participants reported a positive influence on their interest in data visualization (P1, P5, P7, P14). P1 mentioned that ChatGPT ``definitely stimulated sort of this interest in learning further into data visualization [...]'', while P5 expressed ``I definitely think it did... After this course, I would say that I would consider a career in something like this.''

\begin{figure}
  \centering
  \includegraphics[width=\linewidth]{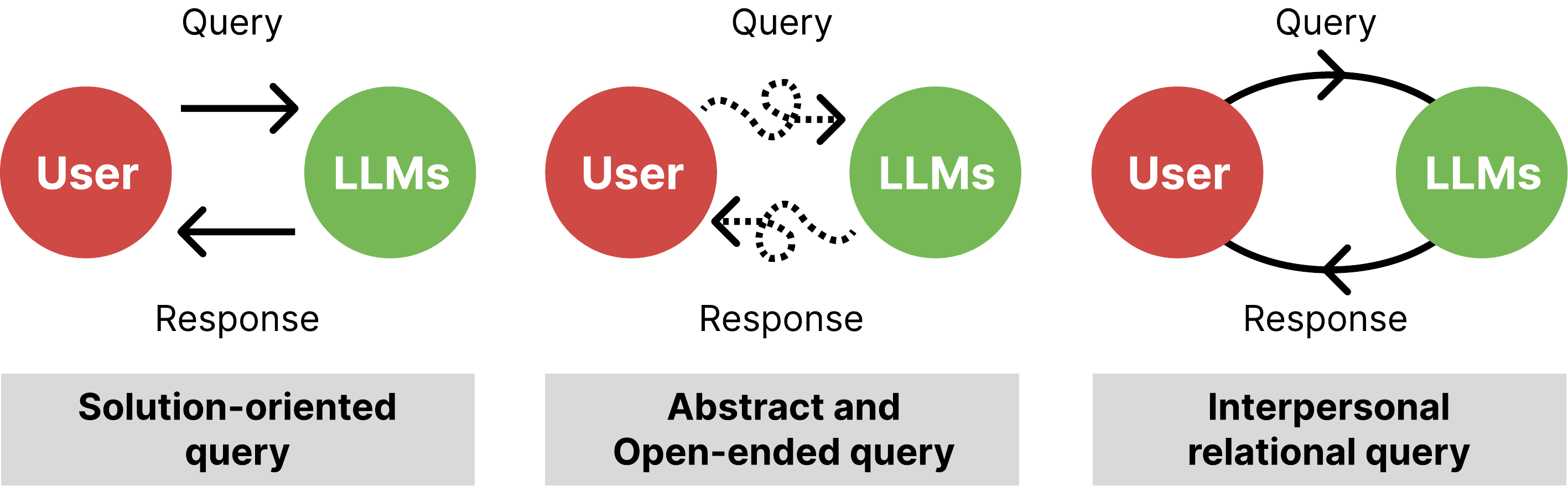}
  \caption{Three themes extracted from analyzing 3,773 queries from 26 users.}
  \label{fig:theme}
\end{figure}

\subsection{Conversation Log Analysis Results}
\label{sec:log-analysis-result}

We identified three categories of user queries: 1) solution-oriented, 2) abstract and open-ended, and 3) interpersonal relational (Figure \ref{fig:theme}). Although we focus on questions asked rather than behavioral characteristics, the first two categories align with Barke et al.'s `acceleration' and `exploration' concepts~\cite{barke2023grounded}, which echo dual-process theories of cognition~\cite{carlston2013oxford}.




A \textbf{solution-oriented} query expects a specific and detailed response with a clear goal. The direction is unambiguous, prompting the system to deliver the desired solution. This is akin to using ChatGPT as a function, where input X (the query) produces output Y (the response). The representative codes include: clarification (215/17,835), implementation skills (190), course content (158), implementation (131), customization (131), where the number in the parenthesis represents their frequency. The example queries in this theme include the following: ``help me create html to have two boxes so that i can put an image and text in each box'', ``i have a csv file with a gender column and a year column I want to add a new column that show the frequency of each gender appears in each years'', ``I am attempting to label points on my scatterplot with this code but it isn't working. What is the problem? \texttt{$<$CODE$>$}'', ``using \texttt{np.isnan} replace \texttt{nan} with \texttt{0}''.

Second, an \textbf{abstract and open-ended} query is one that is broad and often vague, where users are seeking learning, growth, or exploration without a definitive endpoint. 
These queries, sometimes vague due to users' uncertainty about their needs, encourage subjective interpretation and discovery, often resulting in general guidance and exploratory insights.
The representative codes include: data visualization (1257), user interface (418), troubleshooting (246), design knowledge (173), user guidance (162). The example queries in this theme include the following: ``how to make my website prettier: $<$CODE$>$'', ``make the colors brighter'', ``that isn't working'', ``what are some aesthetic symbols i can add to my header.''

Lastly, an \textbf{interpersonal relational} query emphasizes human-like interaction, prioritizing emotional connection and understanding over achieving specific outcomes. The focus is on cultivating empathy, support, or companionship, with users engaging in conversations with ChatGPT as though it were human. The representative codes include: \textit{confusion} (30), \textit{frustration} (29), \textit{expectation} (8), \textit{misunderstanding} (7), \textit{confirmation} (7). The example queries in this theme include the following: ``That certainly did not work'', ``This did not change the outcome at all. Im becoming hopeless im so unsure why it wont work....'', ``Why are these margins not applying! \texttt{$<$CODE$>$} ill tip \$5'', ``thank you. it looks good. i have a parapraph, how can i achieve if i click the word red, the red bar will be highlighted.'', ``sorry i give you the wrong code. review this one: \texttt{$<$CODE$>$}''.

Among 17,835 codes, 53.0\% (9,457) is assigned to solution-oriented query, 46.1\% (8,217) is assigned to abstract and open-ended query, and 0.9\% (161) is assigned to interpersonal relational query.



We categorized participants based on two dominant query types: abstract/open-ended and solution-oriented. 19 participants primarily used solution-oriented queries, while eight favored abstract/open-ended queries. We analyzed theme distribution across different assignment types. Solution-oriented queries outnumbered abstract/open-ended queries for all assignments. Specifically, the proportion of solution-oriented queries was 61.6\% for A1, 64.6\% for A2, 51.3\% for A3, and 55.1\% for A4. Conversely, abstract and open-ended queries comprised 38.4\% for A1, 35.4\% for A2, 48.7\% for A3, and 44.9\% for A4. The largest disparity between the two themes occurred in A2 (a 29.3\% difference), while the smallest was in A3 (a mere 2.5\% difference).

\subsection{Integrated Analysis Results}
When we examined the relationship between the themes of queries used by participants and their final grades, we hypothesized that participants who favored solution-oriented queries might achieve higher grades. However, we did not observe a significant difference (\textit{z}=-0.70, \textit{p}=0.49) between the open-ended query group (grade: \textit{M}=97.9, \textit{SD}=4.76) and the solution-oriented query group (grade: \textit{M}=96.4, \textit{SD}=3.36). 
We also categorized users into two groups based on whether they used interpersonal relational queries. As a result, six users employed interpersonal relational queries, while 20 users did not. The final grade for the group that used these queries was lower (\textit{M}=96.4, \textit{SD}=4.49) than that of the group that did not use them (\textit{M}=97.9, \textit{SD}=4.46); however, this difference was not statistically significant (\textit{z}=0.90, \textit{p}=0.45). 

We analyzed the correlation between students' grades and ChatGPT usage (turn-taking interaction frequency and query character count) across four assignments. For interactions, A1 showed a slight negative correlation ($R^2$=0.008), A2 had no correlation ($R^2$=0.000), and A3 and A4 showed slight positive correlations ($R^2$=0.023 and $R^2$=0.049, respectively). Overall, interactions had minimal impact on grades. In terms of query length, after preprocessing to replace codes, data, and links with placeholders, both A1 and A2 displayed slight negative correlations ($R^2$=0.053 and $R^2$=0.004, respectively), while A3 and A4 showed slight positive correlations ($R^2$=0.013 and $R^2$=0.024, respectively).

We also examined the correlation between students' opinions in the final course survey and their final grades. Although there were no significant relationships, there was a slight negative correlation between the experience ratings and final grades ($\rho=0.20, p=0.33$). Conversely, there was a slight positive correlation between the final grades and time ratings ($\rho=0.14, p=0.50$), engagement ratings ($\rho=0.23, p=0.23$), and quality ratings ($\rho=0.27, p=0.19$). Additionally, we explored the relationship between intermediate survey factors (frequency, confidence, etc.) and student grades, but these also yielded insignificant impacts.


\section{Limitations}
Our work has several limitations: participation was voluntary, possibly introducing selection bias; participants may have had pre-existing favorable or unfavorable views toward ChatGPT; extra credit for participation could have influenced motivation; and the high concentration of grades in the 90th percentile reduces variability, making it challenging to discern differences or achieve statistical significance.

Our study's results may be specific to the ChatGPT version used at the time. Some students may have used the paid version with advanced features like image and dataset reading. Additionally, despite instructions to use the same chat channel for each assignment, some may have started multiple chats, complicating data consistency. 

\section{Discussion}

The unanimous positive reception of ChatGPT underscores its potential for LLM-based chatbots in education. Participants valued its versatility and efficiency in coding tasks and its availability outside the classroom, echoing benefits reported in related work (Section \ref{sec:related-work}) and other professional surveys~\cite{liang2024large,stackoverflow2023survey}. Our results also highlight ChatGPT's benefits in helping students get started and facilitating exploratory learning in a project-oriented data visualization course.

\subsection{Considerations for AI-Enhanced Learning Systems}

Challenges raised by students highlight opportunities for tailored educational solutions, especially for visually oriented courses like data visualization. We discuss design considerations for these systems below.

\paragraph{Encourage creative design inquiries}
Our study indicates that students rarely inquired about data visualization design knowledge, with only a few engaging in basic chart suggestions. Although not critically reflected in their grades, in-class design feedback sessions revealed that students frequently made less effective design choices, such as using continuous scaling for categorical data and inconsistent color schemes across multiple charts. While the capability of ChatGPT or LLMs in visualization knowledge remains largely unexplored, previous work demonstrates that ChatGPT can provide design feedback on creative work~\cite{kim2023good,duan2024generating}. Although the current LLM-chatbot is constrained by its chat interface, future educational systems could offer user interfaces that \textbf{better support design-oriented tasks}.

\paragraph{Support articulation in student queries and response verification}

Students noted that inaccurate responses are costly and can be attributed to many factors. While the limitations of the knowledge base and memories are inherent in LLMs, enhancing the user interface could help students mitigate misleading responses by aiding them in formulating better queries and verifying the quality of LLM responses.
For instance, various prompting strategies can be employed to refine user queries, such as using \textit{prompt chaining} to \textbf{break down tasks into similar subtasks} and employing ReAct prompting~\cite{yao2022react} to \textbf{generate reasoning traces for response transparency}. Furthermore, one can develop an \textbf{external knowledge base~\cite{lewis2020retrieval} about data visualization to augment the LLM} and provide more specific guidance on query formulation and response evaluation.

\paragraph{Support diverse input and output for fluid interactions}

Many frustrations arose from the lack of fluid conversations, often exacerbated by the text-oriented mode of communication, which made it difficult to contextualize participants' design intentions. Although some students who paid for enhanced services could upload images or datasets, there was still a demand for further improvements in input and output capabilities. For example, participants sought file-reading capabilities, such as processing a set of JavaScript, HTML, and CSS files together and \textbf{executing the output for visual confirmation}. Additionally, the ability to \textbf{visually annotate subregions in chart images} could further clarify user intents. Supporting such multimodal interaction in future systems can help address the concern expressed by P7, who stated, ``I didn't know the technical terminology or vernacular to be able to fully describe my intentions.''

\paragraph{Beyond being a mere problem solver}

Many students viewed ChatGPT as a simple tool rather than a learning aid, with P4 noting that it ``just gave me the code'' without facilitating a deeper understanding of ``how it works.'' This highlights a gap in fostering learning and creativity, raising concerns about potential overreliance. Future tailored solutions should offer \textbf{progressive learning opportunities} that emphasize intermediate steps and tailor assistance to students' knowledge levels to overcome incomprehensible responses. Moreover, it would be crucial to define clear boundaries where LLM assistance is beneficial and where students need to develop their own skills, by \textbf{promoting meta-learning strategies} or \textbf{student-in-the-loop learning}~\cite{sellen2023rise}. This balance is essential for an effective learning aid, complementing rather than replacing students' development.
 
\paragraph{Addressing potential equity gaps among students}

We observed concerns that students' ability to effectively use ChatGPT and afford advanced services might widen educational disparities. Some students adeptly extract valuable responses from ChatGPT, while others struggle to formulate effective questions and mistakenly accept inaccurate responses. Likewise, paid users can work more efficiently, benefiting from advanced inputs and outputs. Future systems should address these disparities by \textbf{considering the diverse backgrounds of students} to reduce unintentional biases and discrimination.

\section{Conclusion}
In this work, we present a study investigating how undergraduate students use and perceive ChatGPT in an upper-level data visualization course. The study results outline various benefits and barriers, reaffirming prior studies and uncovering additional issues. Based on the lessons learned, we discuss future opportunities for a more tailored education system that encourages design-oriented tasks and facilitates multimodal interaction and student engagement. For future work, we plan to build such a system with an improved user interface and an augmented knowledge base.

\section{Acknowledgement}
This research was supported by the National Science Foundation under Grant No. 2146868.
\bibliographystyle{plain}
\bibliography{references}

\end{document}